\documentclass[%
 aip,
 amsmath,amssymb,
 reprint,%
]{revtex4-1}

\usepackage{graphicx}
\usepackage{dcolumn}
\usepackage{bm}
\usepackage{xcolor}

\begin{document}


\title[]{Computing the Helmholtz Capacitance of Charged
  Insulator-Electrolyte Interfaces from the Supercell Polarization}

\author{Chao Zhang}
\email{chao.zhang@kemi.uu.se}
\affiliation{Department of Chemistry-\AA ngstr\"om Laboratory, Uppsala
  University, L\"agerhyddsv\"agen 1, BOX 538, 75121, Uppsala, Sweden
}%


\date{\today}

\begin{abstract}
Supercell modelling of an electrical double layer (EDL) at electrified
solid-electrolyte interfaces is a
challenge. The net polarization of EDLs arising from the fixed
chemical composition setup leads to uncompensated EDLs under periodic
boundary condition
and convolutes the calculation of the Helmholtz capacitance  [Zhang
and Sprik, Phys. Rev. B, {\bf 94}, 245309 (2016)].  Here we
provide a new formula based on the supercell polarization at zero
electric field $\bar{E}=0$ to calculate the Helmholtz
capacitance of charged insulator-electrolyte interfaces and validate it using atomistic simulations. Results are
shown to be independent of the supercell size. This formula gives a
shortcut to compute the Helmholtz capacitance without locating the
zero net charge state of EDL and applies directly to any standard
molecular dynamics
code where the electrostatic interactions are treated by the Ewald
summation or its variants.

%
\end{abstract}

\maketitle

%

Charged insulating oxides-electrolyte interfaces are commonly found in
electro/geochemistry~\cite{Westall:1980cols,Trasatti:1996cols,Israelachvili:2011Is}. The charge of insulator surface comes from the
acid-base chemistry. It is negatively charged because of the
deprotonation of the adsorbed water, when pH goes above the point of
zero charge (PZC). On the other hand, it can become positively charged by
protonation when pH goes below PZC~\cite{Trasatti:1996cols}. The charged
insulator surface will naturally polarize surrounding water molecules
and attract counterions from the electrolyte to form the electric
double layer (EDL). The most important quantity to characterize EDL is
its capacitance.

For insulating oxides (or semiconducting oxides at the flatband
condition)~\cite{Nozik:1996gv,Gratzel:2001ub}, the capacitance can be
written as two distinct components connected in series:

\begin{equation}
\label{C_EDL}
1/C_\text{EDL}=1/C_\text{H}+1/C_\text{GC}
\end{equation}

The first component $C_\text{H}$ is the Helmholtz capacitance due to
the chemisorption of hydroxide groups or protons and the attraction
of counterions. The dimension of $C_\text{H}$ is of a molecular size. The second component $C_\text{GC}$ called Gouy-Chapman
capacitance, stems from the diffusive electrolyte and depends on the
ionic strength. Because the diffuse ionic layer has a much higher
capacitance and the inverse $C_\text{GC}$ term turns to be rather small, this makes the Helmholtz
capacitance $C_\textrm{H}$ the leading term (similar to the dead-layer effect
at water interfaces~\cite{Zhang:2018tl}) and the focus of this study. 

\begin{figure} [h]
\includegraphics[width=0.95\columnwidth]{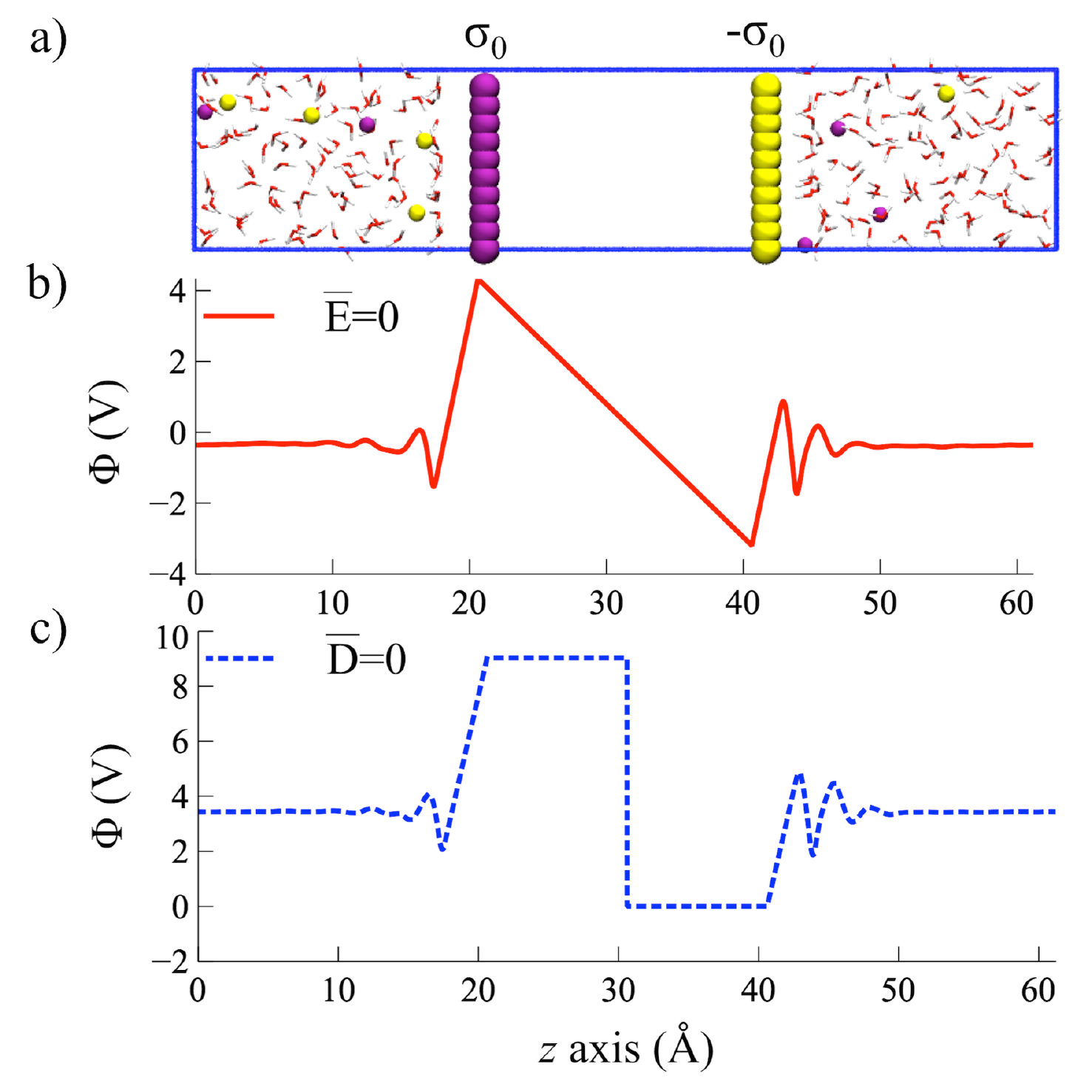}
\caption{\label{elpt} a) Periodic model of two complementary charged
  insulator-electrolyte interfaces used as the model system in this
  study. The charged insulator is modelled as a pair of rigid atomic
  walls with opposite charge separated by a vacuum region (the
  insulator). The surface charge is uniformly distributed over area
  $A$ with a charge density $\sigma_0$. Positive charges are in purple and negative
  charges are in yellow. b)  The electrostatic potential profile
  $\Phi(z)$ averaged over the perpendicular $x$ and $y$ directions  at
  $\bar{E}=0$ and $A\sigma_0=2e$; c) The electrostatic potential profile
  $\Phi(z)$ averaged over the perpendicular $x$ and $y$ directions at
  $\bar{D}=0$ and $A\sigma_0=2e$. }
\end{figure}

Computing $C_\textrm{H}$ may not be as easy as it seems. Under periodic boundary
condition (PBC), two insulator-electrolyte interfaces can be
charged up either symmetrically (same amounts and types of surface charges) ~\cite{Cheng:2014eh, Dewan:2014ig,
  Sultan:2014hj, Parez:2014dx, Hocine:2016da, Sulpizi:2016scirep} or asymmetrically
(same amounts but opposite types of
surface charges)~\cite{GuerreroGarcia:2013jq, PfeifferLaplaud:2016cg}. However,
only in the asymmetric setup, the chemical
composition can be kept fixed at different surface charge
densities, which satisfies the actual
experimental conditions. In the asymmetric setup (Fig.~\ref{elpt}a), supercell contains two parallel EDLs and a net polarization. As
a consequence, each EDL is not fully compensated under PBC. This can
be easily inferred from the electrostatic potential profile of the
model system (Fig.~\ref{elpt}b), where there is an electric field in the insulator
region (Here we simply used vacuum for the proof-of-concept). According to Gauss's theorem, a finite field means the enclosed
body (an EDL for this case) bears a net charge. This net charge in
EDLs is the manifestation of a finite-size error which plagues the
computation of the Helmholtz capacitance. 

Built on finite field methods developed
by Stengel, Spaldin and Vanderbilt (SSV)~\cite{Stengel:2009cd,
  Stengel:2009prb} for ferroelectric systems and extended later to finite-temperature
simulations~\cite{Zhang:2015ms, Zhang:2016ho}, we have proposed and
validated two methods
to compute the size-independent Helmholtz capacitance of EDLs of charged
insulator-electrolyte interfaces under PBC~\cite{edl2016}. The
first one is based on constant electric field $\bar{E}$
simulations. By locating the zero net charge (ZNC) state of EDL, the
corresponding external field $\bar{E}$ gives directly the
Helmholtz capacitance of EDLs~\cite{edl2016}. Subsequently, this method was extended
to study charge compensation between polar surfaces and electrolyte
solution~\cite{Sayer:2017cw}. The second one is based on constant
electric displacement $\bar{D}$ simulations. The differential of the
itinerant polarization with respect to the imposed surface charge density at constant $\bar{D}$ gives an efficient estimation of the overall Helmholtz capacitance of EDLs~\cite{edl2016}.

These two methods were devised from our analysis of 
a Stern-like model as the continuum counterpart of the atomistic
system. In the second method based on constant $\bar{D}$ simulations,
one gets the Helmholtz capacitance $C_\textrm{H}$ without
locating ZNC state of EDL~\cite{edl2016}. This suggests that it should be possible to
derive the corresponding formula without relying on the Stern-like
continuum model. In this Letter, we rederive the method for calculating the Helmholtz capacitance at
constant $\bar{D}$ and show that this leads to a new formula to compute the
Helmholtz capacitance using the supercell polarization at 
$\bar{E}=0$ (i.e. the standard Ewald boundary condition) through thermodynamics relations. This new formula is then verified by molecular
dynamics (MD) simulation based on a simple point-charge (SPC)-like model of the charged
insulator-electrolyte system. The resulting Helmholtz
capacitance is shown to be independent of the supercell size and in
excellent agreement with that obtained from constant electric displacement
$\bar{D}$ simulations~\cite{edl2016}.

What we start with is the hybrid SSV constant $\bar{D}$ Hamiltonian,
which can be derived either from the thermodynamics
argument originally~\cite{Stengel:2009cd} or from a current dependent
Lagrangian as shown recently~\cite{Sprik:2018kr}:
\begin{equation}
\label{uvdb}
H_D\left(v ,\bar{D} \right) = H_{\mathrm{PBC}}(v)+ \frac{\mathit{\Omega}}{8 \pi}
  \left(\bar{D} - 4 \pi P(v) \right)^2
\end{equation}
where $P$ is the itinerant polarization in the direction of
$\bar{D}$ (See
Secs. IV B and IV C in Ref. ~\cite{edl2016} for the elaboration), which is formally
defined as a time integral of the volume integral of current ~\cite{King-Smith:1993prb,Resta:1994rmp,Resta:2007ch,
  Caillol:1989jcpa,Caillol:1989jcpb,Caillol:1994ho}. $\Omega$ is the supercell
volume and $v= (\mathbf{r}^N ,\mathbf{p}^N)$ stands for the collective
momenta and position coordinates of the $N$ particles in the
system. The bar over $D$ emphasizes that it is a variable instead of
an observable. ``Hybrid'' means the field is only applied in the
direction perpendicular to the surface. 

The extended Hamiltonian $H_D\left(v, \bar{D} \right)$ of Eq.~\ref{uvdb} generates a field dependent partition function
\begin{equation}
\label{zfmd}
Z_D (\bar{D})= \int d\nu \exp \lbrack - \beta  H_D\left(v ,\bar{D} \right) \rbrack
\end{equation}
 $\beta = 1/k_{\mathrm{B}}T $ is the inverse temperature. The combinatorial prefactor $1/(h^{3N} N!)$ has been omitted.

The expectation value of an observable $X$ is 

\begin{equation}
 \langle X \rangle =\int d\nu \frac{ X \exp \left[- \beta H_D \left( \nu, \bar{D} \right) \right]}{Z_D(\bar{D})}
\end{equation}

The electric displacement $\bar{D}$ is related to the electric field $E$
according to the definition:
\begin{equation}
\label{ddef}
\bar{D}= E+4\pi P
\end{equation}

This leads to the expectation value of the voltage difference $\Delta V$
crossing the supercell as:
\begin{equation}
\langle \Delta V \rangle = -\langle E \rangle L =
-(\bar{D}-4\pi\langle P \rangle)L
\end{equation}
where $L$ is the dimension of the supercell in the $z$ direction
which is along the surface normal.

Then, the overall capacitance according to the definition is:

\begin{eqnarray}
C_\textrm{tot}=\frac{C_H}{2}&=& \left (\frac{\partial \sigma_0}{\partial \langle
    \Delta V \rangle } \right )_{\bar{D}}  \\
&=& \left (\frac{\partial \sigma_0}{-\partial(\bar{D}-4\pi\langle P
    \rangle)L } \right )_{\bar{D}} \\
&=& \frac{1}{4\pi L}\left (\frac{\partial \sigma_0}{\partial \langle P \rangle}
\right )_{\bar{D}}
\label{CH_dPD}
\end{eqnarray}

Here we assume again that two EDLs connected in series have the same
Helmholtz capacitance $C_\textrm{H}$ (Fig.~\ref{elpt}a). In other
words, $C_\textrm{H}$ is the average Helmholtz capacitance at a
surface charge
density $|\sigma_0|$. We notice that Eq.~\ref{CH_dPD} is the same differential formula for the
capacitance of the Helmholtz capacitance $C_\textrm{H}/2$ at
constant $\bar{D}$, as derived from the linear electric equation of
state using the Stern-like continuum model in our previous work~\cite{edl2016}. 

Because $\bar{D}$ and $\bar{E}$ are thermodynamic conjugate variables,
this allows us to find out the corresponding relation of Eq.~\ref{CH_dPD} at
$\bar{E}$. The procedure we took is similar to that used to establish the thermodynamic relation between heat capacities at constant volume and at constant pressure.

First, we introduce following two expressions:

\begin{eqnarray}
\left (\frac{\partial \sigma_0}{\partial \bar{D}}  \right )_{P} &=&
-\left (\frac{\partial \sigma_0}{\partial P} \right )_{\bar{D}} \left
  (\frac{\partial P}{\partial \bar{D}} \right )_{\sigma_0}  \\
\left (\frac{\partial \sigma_0}{\partial \bar{E}}  \right )_{P} &=&
-\left (\frac{\partial \sigma_0}{\partial P} \right )_{\bar{E}} \left
  (\frac{\partial P}{\partial \bar{E}} \right )_{\sigma_0}  
\end{eqnarray}

The ratio between them leads to:
\begin{eqnarray}
\left (\frac{\partial \sigma_0}{\partial P} \right
  )_{\bar{D}} \left (\frac{\partial P}{\partial \sigma_0} \right
  )_{\bar{E}} &=& \left (\frac{\partial
                  \bar{D}}{\partial \bar{E}} \right )_{\sigma_0} \left (\frac{\partial \bar{E}}{\partial \bar{D}}
                  \right )_{P}\\
&=& \epsilon_\perp \left (\frac{\partial \bar{E}}{\partial \bar{D}}
    \right )_{P}
\label{partial1}
\end{eqnarray}

Here $\epsilon_\perp$ is the overall dielectric constant of
the heterogenous system in the direction perpendicular to the surface
and the subscript $\sigma_0$ of $\epsilon_\perp$ is omitted. 

Then, the second term on the right hand side of Eq.~\ref{partial1}
can be rewritten as,

\begin{eqnarray}
\left (\frac{\partial \bar{E}}{\partial \bar{D}}\right )_{P} &=& \left
  (\frac{\partial \bar{E}}{\partial \sigma_0}\right )_{P} \left
                                                                 (\frac{\partial \sigma_0}{\partial \bar{D}}\right )_{P} \\
&=& \left
  (\frac{\partial \bar{E}}{\partial \sigma_0}\right )_{P} \left
                                                                 (\frac{\partial
    \sigma_0}{\partial (\bar{E}+4\pi P)}\right )_{P} \\ 
&=& \left
  (\frac{\partial \bar{E}}{\partial \sigma_0}\right )_{P} \left
                                                                 (\frac{\partial
    \sigma_0}{\partial \bar{E}}\right )_{P} \\
&=& 1
\label{partial2}
\end{eqnarray}

Combining Eq.~\ref{partial1} and Eq.~\ref{partial2}, we obtain a
key intermediate result:

\begin{equation}
\left (\frac{\partial \sigma_0}{\partial P} \right
  )_{\bar{D}} \left (\frac{\partial P}{\partial \sigma_0} \right
  )_{\bar{E}} =\epsilon_\perp 
\label{dPD2}
\end{equation}

Inserting Eq.~\ref{dPD2} into Eq.~\ref{CH_dPD}, one ends up with the desired relation:

\begin{equation}
\frac{C_H}{2}=\frac{\epsilon_\perp}{4\pi L}\left (\frac{\partial \sigma_0}{\partial \langle P \rangle}
\right )_{\bar{E}}
\label{CH_dPE}
\end{equation}

This is the corresponding differential formula for the overal Helmholtz
capacitance at constant $\bar{E}$.

For the system at $\bar{E}=0$ and under PBC, it is known from the
linear response theory that~\cite{Neumann:1983kg}:

\begin{equation}
\label{LRT}
\epsilon_\perp = \left (\frac{ \partial \langle P \rangle }{\partial
    \bar{E}}\right)_{\bar{E}=0} +1 = 4\pi\beta \mathit{\Omega} \left(\langle P^2\rangle_{\bar{E}=0} -
  \langle P\rangle^2_{\bar{E}=0} \right) +1
\end{equation}

Since $\langle P \rangle =0$ for $\sigma_0 = 0$, therefore, the equation for computing $C_\textrm{H}$ is simply:
\begin{eqnarray}
\label{CH_fluc0}
\frac{C_\textrm{H}}{2} &=& \frac{\epsilon_\perp}{4\pi
                           L}\frac{\sigma_0}{\langle P
                           \rangle}_{\bar{E}=0} \\
&=&\frac{\sigma_0 \left[ 4\pi\beta \mathit{\Omega} \left(\langle
                          P^2\rangle_{\bar{E}=0} - \langle
                          P\rangle^2_{\bar{E}=0} \right) +1 \right] }{4\pi L \langle P
  \rangle_{\bar{E}=0} } 
\label{CH_fluct}
\end{eqnarray}

Eq.~\ref{CH_fluct} is the main result of this work, where the polarization
fluctuation is a necessary piece of information for computing the Helmholtz
capacitance at $\bar{E}=0$, i.e. the standard Ewald boundary
condition, for the generic system showed in Fig.~\ref{elpt}a.

To test whether this formula gives a size-independent
estimator of the Helmholtz capacitance, we have performed MD simulation of a SPC-like model, which is  familiar
from many studies of electrode-electrolyte interfaces~\cite{Spohr:1999tm,Dimitrov:2000ti,Fedorov:2008gh,Zarzycki:2010kja,LyndenBell:2012iq,
  Sultan:2014hj, Parez:2014dx, Dewan:2014ig, Hocine:2016da}. The
electrolyte consists of 202 water molecules, 5 Na$^+$ and 5 Cl$^-$
ions. The oppositely charged insulator slab was modelled as two rigid
uniformly charged atomic walls plus a vacuum slab in between as the
insulator. The simulation box is rectangular. The length in $x$ and
$y$ direction is 12.75~\AA~ and the length in $z$ direction varies
from 61.24~\AA~to 121.24~\AA~depending on the thickness of the
insulator (vacuum in this case). Water are described by the SPC/E model
potential~\cite{Berendsen:1987uu} and alkali metal ions are modelled as point
charge plus Lennard-Jones potential using the parameters from
Jung and Cheatham~\cite{Joung2008, Zhang:2010zh}. The van der
Waals parameters of the particle in the rigid wall were simply chosen
to be the same as those of oxygen atom. The MD
integration time step is 2~fs and trajectories were accumulated for
10ns for each combination of the charge density and the electric
boundary condition. The electrostatics was computed using Particle
Mesh Ewald (PME) scheme~\cite{Ewald}. Short-range cutoffs for the Van
der Waals and Coulomb interaction in direct space are 6~\AA. The
temperature was controlled by a Nos\'e-Hoover chain thermostat set at
298K~\cite{martyna92}. These technical setting are the same as in the
previous work~\cite{edl2016} and all simulations were done with a
modified version of GROMACS 4 package~\cite{Hess2008}. In the case of
$\bar{D}=0$ simulation, we used the hybrid constant $\bar{D}$
Hamilton shown in Eq.~\ref{uvdb}. This implies a static and homogenous
$\bar{D}$ field was only applied in the direction perpendicular to the surface
(i.e. $z$ direction) over the whole simulation box. Regarding the itinerant polarization $P$, it differs from
the conventional cell polarization $P^{\textrm{cell}}(t) =  \frac{1}{\Omega} 
\sum_i^{\textrm{cell}} q_i \textrm{nint}(L^{-1}z_i(t))$ by preserving
the continuity of time-integrated current~\cite{King-Smith:1993prb,Resta:1994rmp,Resta:2007ch,
  Caillol:1989jcpa,Caillol:1989jcpb,Caillol:1994ho}. This means that
the iternative polarization $P$ is continuous throughout the trajectory
and particles need
to be tracked from t=0 if they leave the MD supercell when computing
the polarization. From the iternative
polariztion $P$, one can also compute the overal dielectric constant
$\epsilon_{\perp}$ following Eq.~\ref{LRT} straightforwardly.

\begin{figure} [h]
\includegraphics[width=0.95\columnwidth]{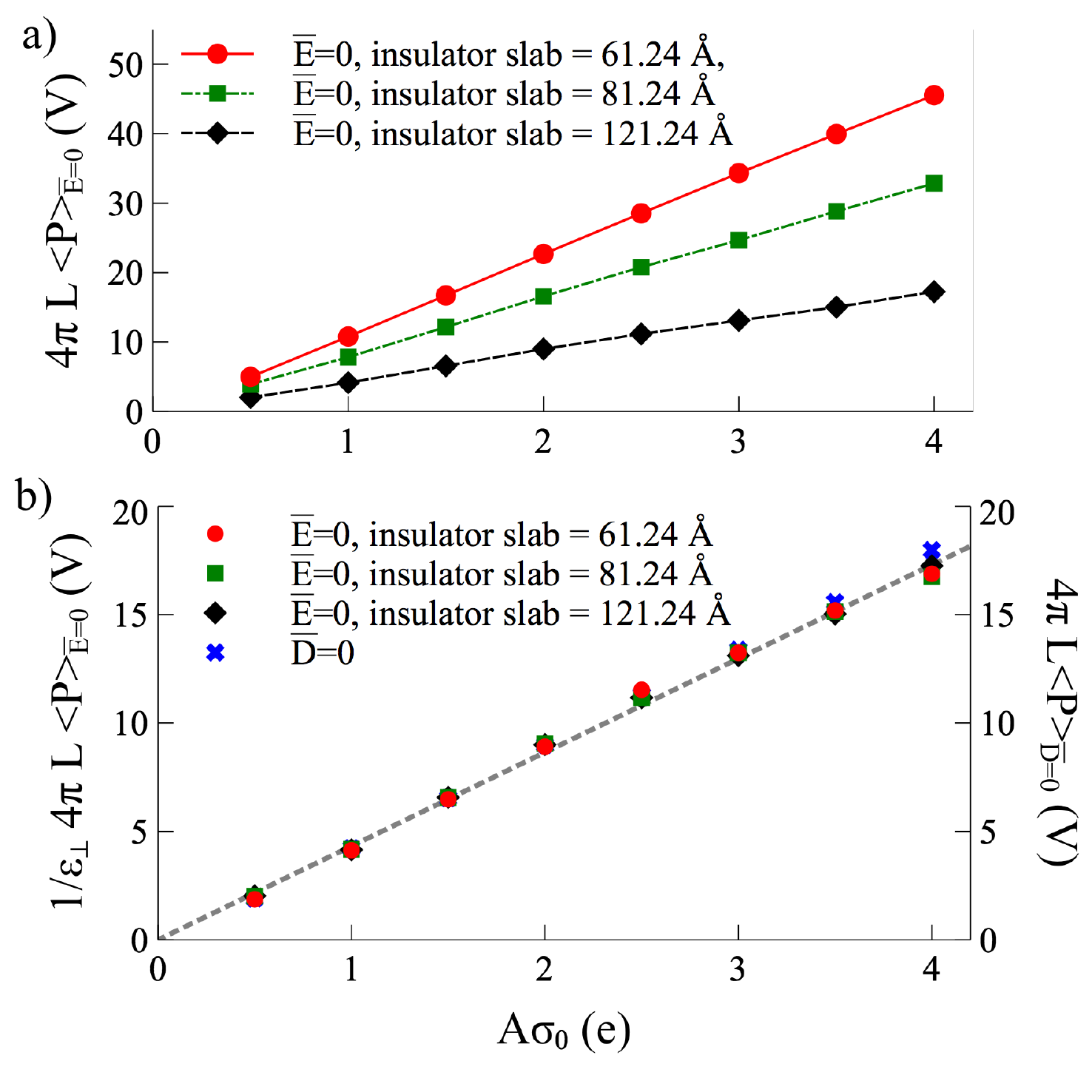}
\caption{\label{pol} a) The polarization potential $4\pi L \langle P \rangle$
  (in volt) as a function of the imposed
  surface charge $A\sigma_0$ of the charged
  insulator-electrolyte system for three different insulator slab sizes
  at $\bar{E}$=0. $L$ is the the box length in $z$ direction,
  perpendicular to the surface; b) The overall dielectric constant $\epsilon_\perp$ weighted polarization potential $1/\epsilon_\perp 4\pi L \langle P \rangle$ (in volt) as a function of the imposed
  surface charge $A\sigma_0$ for the same system at
  $\bar{E}=0$. This should be compared to the polarization potential $4\pi L \langle
  P \rangle$ (in volt) as a function of the imposed
  surface charge density $A\sigma_0$ at $\bar{D}=0$. $\epsilon_\perp$ was
  computed according to Eq.~\ref{LRT} for the system of different surface charge density and insulator slab size.}
\end{figure}

The polarization potential $4\pi L \langle P \rangle$ has the same unit
as the voltage and that is what we plotted in Fig.~\ref{pol}a. As shown
in the Figure, the polarization potential at $\bar{E}=0$ has a linear
relation with respect to the imposed charge density $\sigma_0$. The slope which is directly related
to the Helmholtz capacitance has a strong size dependence of the
supercell. This confirms that the insulator also
contributes to the total capacitance because of the existing field in
the insulator region under PBC (Fig.~\ref{elpt}b). This is the
finite-size error that we want to remove.

Following Eq.~\ref{CH_fluc0}, we weighted the 
polarization potential $4\pi L \langle P \rangle$ at $\bar{E}=0 $ by the overall dielectric
constant $\epsilon_\perp$ and results are shown in Fig.~\ref{pol}b. As seen in the
Figure, data points for difference sizes of supercell at the same charge
density $\sigma_0$ superimpose with each other.  By fitting these
data to a linear function passing the origin,  one can obtain the slope which gives the
inverse of the Helmholtz capacitance. To check the consistency, we also
computed the polarization potential $4\pi L \langle P \rangle$ at
$\bar{D}=0$ as the reference (Fig.~\ref{elpt}c). One needs to pay
attention that the D value
which restores the ZNC state of EDL for the insulator centered
supercell is subject to the modulation of the polarization quantum
$4\pi e/A$, i.e. $D^{n}_\text{ZNC}=n4e \pi/A$ ~\cite{edl2016} where $n$
is an integer. For the
supercell shown in  Fig.~\ref{pol}a with $A\sigma_0 = 2e$, $D^{n}_\text{ZNC}=0$.

As shown in Fig.~\ref{pol}b, the polarization
potential $4\pi L \langle P \rangle$ at $\bar{D}=0$ at the same
charge density are spot on the weighted
polarization potential $1/\epsilon_\perp 4\pi L \langle P \rangle$ at $\bar{E}=0$. This
suggests both Eq.~\ref{CH_dPE} and Eq.~\ref{CH_dPD} give the same
result for the Helmholtz capacitance, which is independent of the
the system size.

In our previous work~\cite{edl2016}, it was demonstrated that a finite
$\bar{E}$ field can be applied to cancel out the existing field in the insulator region and to restore
the point of ZNC of EDLs. Subsequently, the Helmholtz capacitance can be
obtained from the value of the restoring field at ZNC as~\cite{edl2016}:

\begin{equation}
\label{CH_pzc}
V_\textrm{znc}=-L\bar{E}_{\textrm{znc}}=2\sigma_0/C_{\textrm{H}}
\end{equation}

Putting Eq.~\ref{CH_pzc} and Eq.~\ref{CH_fluct} together, we obtain a new
estimator of the external potential needed to restore ZNC state just
using the supercell polarization at zero electric field:
\begin{equation}
\label{pzc_fluc}
V_\textrm{znc}=-L\bar{E}_{\textrm{znc}} = \frac{4\pi L \langle P
  \rangle_{\bar{E}=0}}{4\pi \beta \mathit{\Omega} \left(\langle P^2\rangle_{\bar{E}=0} - \langle P\rangle^2_{\bar{E}=0} \right) +1 } 
\end{equation}

For the surface charge $A\sigma_0 = 2.0e$, the above formula gives an
estimate of $V_\textrm{znc}$ as 9.0 V. This value should be
compared to 8.9 V as reported previously for the same SPC-like system
by monitoring the net charge of EDL $Q_\textrm{net}$ as a function of
the applied voltage $V_\textrm{ext}$~\cite{edl2016}. Therefore,
Eq.~\ref{pzc_fluc} is also validated.

 Like its constant $\bar{D}$ variant in Eq.~\ref{CH_dPD}, Eq.~\ref{CH_fluct} does not require an \emph{additional} vacuum slab in the first
place, which is a relief for plane-wave based electronic structure
calculation.  Here, the main advantage of using this formula
to compute the Helmholtz capacitance is that it works
directly with any standard MD code in which the electrostatic interactions are
treated by the Ewald summation (or its variants). This was achieved by introducing the overal dielectric
constant $\epsilon_\perp$ which absorbs the finite-size effect. Thus,
it would be interesting in future works to look closer at the role of $\epsilon_\perp$ in supercell modeling of heterogenous
systems. Nevertheless, it is worth to mention that
Eq.~\ref{CH_fluct} only provides a shortcut to compute the Helmholtz
capacitance and a finite field (either $\bar{E}$ or $\bar{D}$) is
still required to restore the ZNC state of EDL in supercell modeling of charged
insulator-electrolyte interfaces.  

\begin{figure} [h]
\includegraphics[width=0.90\columnwidth]{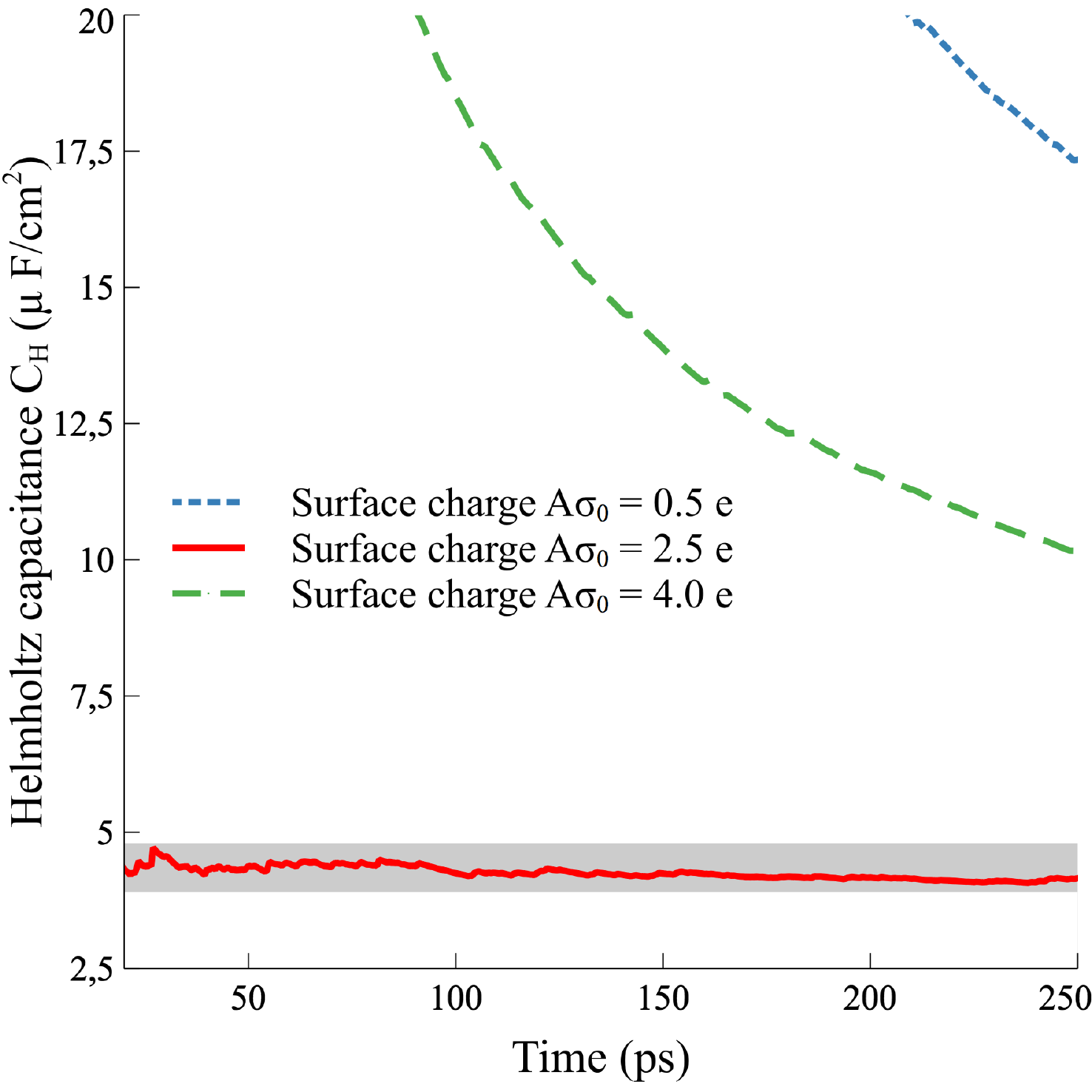}
\caption{\label{convergence} The running average of the Helmholtz
  capacitance $C_\text{H}$ calculated from the supercell polarization
  using Eq.~\ref{CH_fluct} at difference surface charges with the
  smallest box length used in this work (i.e. $L=61.24$~\AA). All simulations were done using the same initial
  configuration extracted from an equilibrated system at surface
  charge $A\sigma_0=2.0~e$ and with the same chemical composition. The
  shaded area indicates $\pm 10$\% deviations from the supposed $C_\text{H}$
  value of this SPC-like model. }
\end{figure}

Before closing this Letter, it is necessary to discuss the
convergence of the Helmholtz capacitance computed from the supercell
polarization. According to the classical Debye theory, switching the
electric boundary condition from constant $\bar{E}$ to constant
$\bar{D}$ would lead to a speed-up of the relaxation time of the
macroscopic polarization by a factor comparable to the dielectric
constant of the medium. This was indeed seen in the simulation of bulk
liquid water~\cite{Zhang:2015ms}. As a consequence, the convergence of
$C_\text{H}$ of charged solid-liquid interfaces can be achieved within 50 ps by using constant $\bar{D}$
simulations (i.e. Eq.~\ref{CH_dPD}) and a SPC-like model (See Fig.~11 in
Ref.~\cite{edl2016}). Instead, Eq.~\ref{CH_fluct} uses the standard
Ewald boundary condition ($\bar{E}=0$) and 
relies on the overal dielectric constant $\epsilon_\perp$ which can have the same
notoriously slow convergence (few nanoseconds) as what we knew for polar liquids (See
Ref.~\cite{Zhang:2016ho} and reference therein). However,
the convergence Eq.~\ref{CH_fluct} of can be achieved within tens of
picoseconds \emph{in practice} if the system was equilibrated at a chosen
surface charge nearby the target value (Fig.~\ref{convergence}). This
leverages the feasibility of applying Eq.~\ref{CH_fluct} in density functional theory based MD
simulations. 

\begin{acknowledgments}
The author thanks M. Sprik for many stimulating discussions and Uppsala University for the support of a start-up grant.
\end{acknowledgments}


%

\end{document}